\documentclass[twocolumn,prl,floatfix]{revtex4}
\usepackage{graphicx,color}
\usepackage{subfigure}
\usepackage{verbatim}
\usepackage{ifpdf}
\usepackage{ifthen} 
\usepackage{url}

\ifpdf
\usepackage[pdftex,colorlinks=true]{hyperref}

\else
\usepackage[dvips,colorlinks=false]{hyperref}

\fi

\usepackage{breakurl}

\hypersetup{
pdftitle={Challenges in continuum modeling of intergranular fracture},
pdfauthor={Valerie R. Coffman, James P. Sethna, Gerd Heber, Mu Liu, Anthony
Ingraffea, Nicholas P. Bailey, Erin Iesulauro Barker},
pdfkeywords={grain boundaries, molecular dynamics, fracture mechanics, finite
  elements, solid state physics,
  computational physics},
bookmarksnumbered,
pdfstartview={FitH},
pdfpagemode=UseThumbs,
breaklinks=true
}

\newcommand{\mycaption}[2][]{
  \ifthenelse{\equal{}{#1}} {\caption{#2}} {\caption[#1]{ {\bf#1} #2}}}

\begin{document}

\title{Challenges in continuum modeling of intergranular fracture}

\author{Valerie R. Coffman}
\thanks{Corresponding author}
\email{valerie.coffman@nist.gov}
\altaffiliation[Current affiliation: ]{Information Technology Laboratory,
National Institute of Standards and Technology, Gaithersburg, MD, 20899}
\affiliation{Laboratory of Atomic and Solid State Physics (LASSP), Clark Hall,
Cornell University, Ithaca, NY 14853-2501, USA}
\author{James P. Sethna}
\affiliation{Laboratory of Atomic and Solid State Physics (LASSP), Clark Hall,
Cornell University, Ithaca, NY 14853-2501, USA}
\author{Gerd Heber, Mu Liu, Anthony Ingraffea}
\affiliation{Cornell Fracture Group, Rhodes Hall,
Cornell University, Ithaca, NY 14853-2501, USA}
\author{Nicholas P. Bailey}
\affiliation{Department of Mathematics and Physics (IMFUFA), DNRF Center ``Glass
and Time'', Roskilde University, P.O. Box 260, DK-4000 Roskilde, Denmark}
\author{Erin Iesulauro Barker}
\affiliation{Los Alamos National Laboratory, Los Alamos, New Mexico 87545, USA}

\date{\today}

\maketitle

\textbf{Intergranular fracture in polycrystals is often simulated by finite elements
coupled to a cohesive-zone model for the interfaces, requiring cohesive laws for
grain boundaries as a function of their geometry. We discuss three challenges in
understanding intergranular fracture in polycrystals.  First, 3D grain boundary
geometries comprise a five dimensional space.  Second, the energy and peak
stress of grain boundaries have singularities for all commensurate grain
boundaries, especially those with short repeat distances. Thirdly, fracture
nucleation and growth depends not only upon the properties of grain boundaries,
but in crucial ways on edges, corners, and triple junctions of even greater
geometrical complexity. To address the first two challenges, we explore the
physical underpinnings for creating functional forms to capture the heirarchical
commensurability structure in the grain boundary properties. To address the last
challenge, we demonstrate a method for atomistically extracting the fracture
properties of geometrically complex local regions on the fly from within a
finite element simulation.}

The nucleation and propagation of cracks in practical engineering materials
depends strongly on the mesoscopic structure; grain boundaries, polyphase
inclusions, dislocations and other defects determine the toughness. Can
continuum computational modeling be used to quantitatively study such
complex failure modes? 

Consider brittle intergranular fracture---rupture at the boundary between two
crystallites. Ignore for the moment problems like embrittlement caused by
impurity segregation to grain boundaries, and assume a clean, single-phase,
equilibrium grain boundary. Direct atomistic
simulations are infeasible for anything larger than nanocrystals; even
simulations that focus on the boundaries~\cite{Quasicontinuum} will be
overwhelmed by the number of relevant atoms for systems larger than microns
in scale. Hence let us imagine a finite-element simulation of the
polycrystal coupled to, say, cohesive-zone models for each 
interface~\cite{needleman-interfaceCZM, xu-needleman, camacho-ortiz,
tvergaard-hutchinson}. (A cohesive law gives the crack opening 
as a function of the traction across the interface; it is often 
parameterized
by a peak stress and a total energy associated with cleaving~\cite{cube-in-cube}.) 
Can one use atomistic models to measure the 
fracture properties of the individual boundaries, and then use these properties
in a realistic continuum fracture simulation?
We outline here some serious challenges involved in continuum modeling of 
intergranular fracture; complete details are available in longer 
publications~\cite{grain-boundary-geo,cube-in-cube}.

The first challenge is that of {\em geometrical complexity}. 
The cohesive law will
depend on the structure of the grain boundary. The macroscopic geometry of
a 3D grain boundary depends on five parameters that describe the
relative orientations of the two grains. 
  (The fracture dynamics may in principle depend on 
  properties of the crack that are not treated explicitly by the cohesive
  zone model, such as the orientation of the crack front
  within the grain boundary or the three separate stress intensity factors.)
The atomistic structure also depends on how the two crystal lattices are 
translated with respect to one another along the three directions, 
which can greatly affect the pattern of atoms along the boundary and hence
the peak stress and energy~\cite{grain-boundary-geo}. One particular shift will
constitute a global energy minimum corresponding to the most natural
configuration~\cite{grain-boundary-geo}. 

In a polycrystal, one grain will have to find an energy minimizing configuration
with several other, neighboring grains. For a particular grain boundary in a
polycrystal, where each grain has been pinned by other neighboring grains, there
will be a competition between elastic straining and plastic deforming. For thick
grains, in equilibrium, and away from intersections, one can show that it is
advantageous for the crystallites to strain slightly to allow the boundary to
find the global energy minimum.

\begin{figure}[thb]
\begin{center}
\includegraphics[width=8cm]{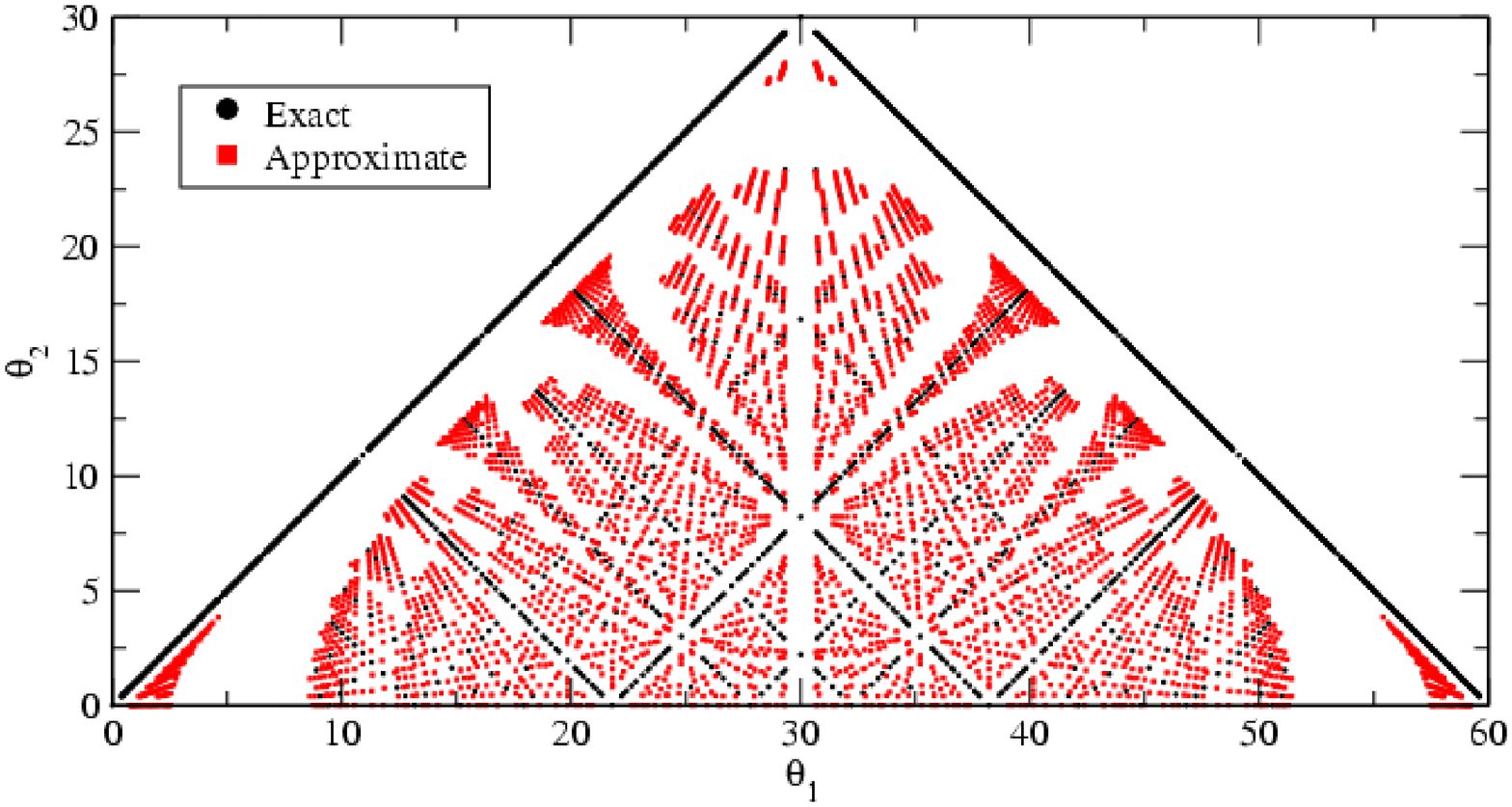}
\end{center}
\mycaption[Commensurate 2D Grain Boundaries.]{The set of points above represent all
2D grain boundary geometries that can be simulated in a periodic box of 70
lattice constants
or less, with a strain of 0.05\% or less.  $\theta_1$ and $\theta_2$ are the tilt
angles that define the grain boundary geometry.  
There are gaps near perfect
crystals, symmetric grain boundaries, and high symmetry grain boundaries because
creating a new, nearby geometry requires adding flaws (extra dislocations) at 
large separations.~\cite{grain-boundary-geo}}
\label{fig:2DCommensurability}
\end{figure}

\begin{figure}[thb]
\begin{center}
\includegraphics[width=7.8cm]{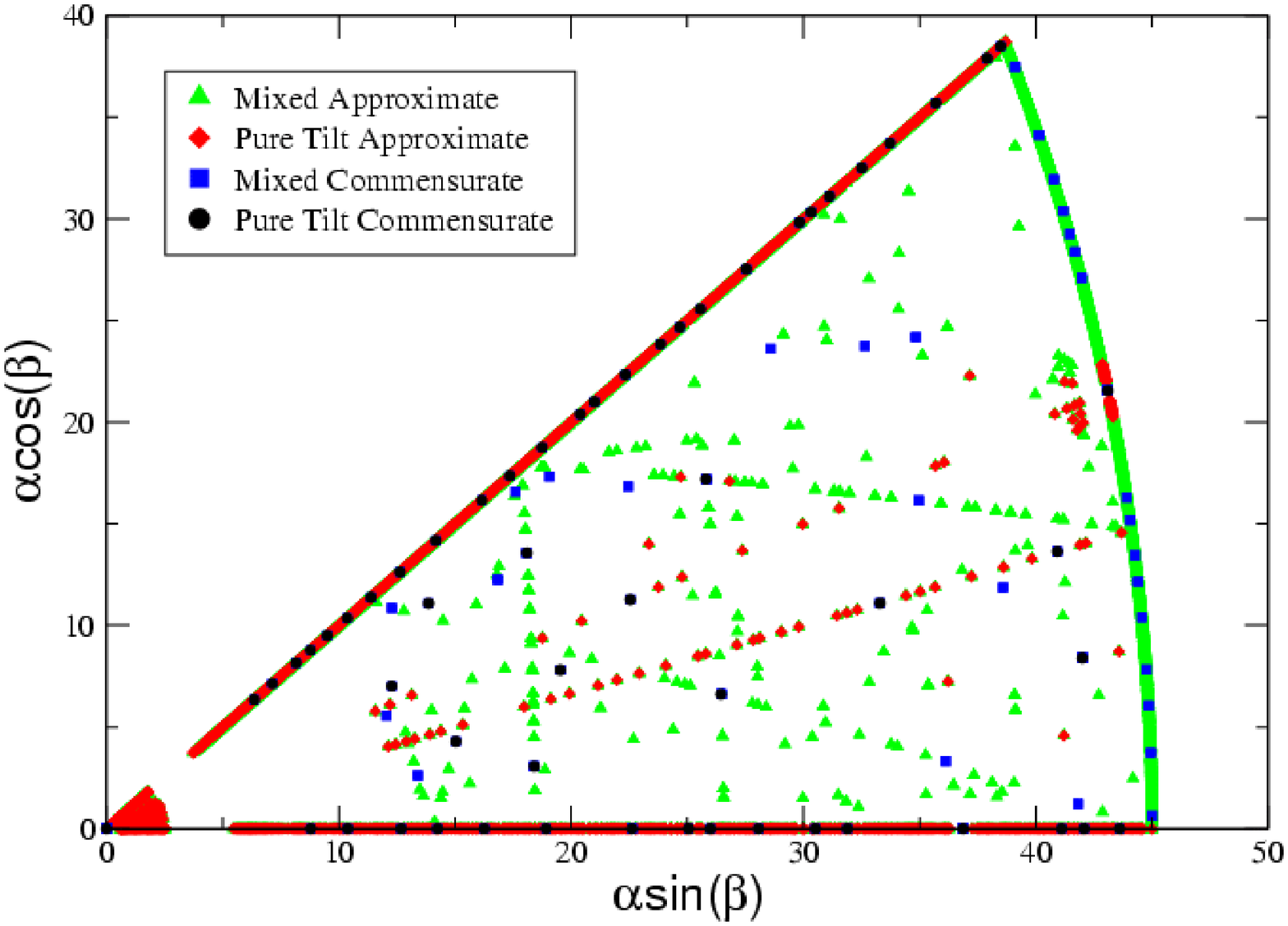}
\end{center}
\mycaption[Commensurate 3D Grain Boundaries.]{The set of points above represent a stereographic
projection (defined for normal vector $\vec{n}$ 
by $\vec{n} = (i,j,k)/ \sqrt{i^2+j^2+k^2} = (\sin \alpha \cos \beta, \sin \alpha
\sin \beta, \cos \alpha)$)
of all FCC surface
orientations that are commensurate with the [100] surface within an area less
than 100 square lattice constants with a strain of .1\% or less.  This
constitutes a three dimensional cross section of the five dimensional space of
grain boundary geometries, where the twist angle parameter has been collapsed
onto a 2D plot.}
\label{fig:3DCommensurability}
\end{figure}

To compute the cohesive properties of grain boundaries efficiently, it is useful
to use periodic boundary conditions in directions perpendicular to the grain
boundary, which demands that the the two crystals have finite repeat distances
along the interface, and that the repeat distances be commensurate with one
another.  We have found a systematic method of finding commensurate grain
boundaries~\cite{grain-boundary-geo}, and have also generalized it to allow for
slight elastic strains to mesh the two crystal boundaries together.  We can
approximate commensurate grain boundaries by allowing small strains in either
direction.  Fig.~\ref{fig:2DCommensurability} describes the commensurate grain
boundaries for 2D hexagonal crystals; Fig.~\ref{fig:3DCommensurability} shows a
cross section of the five dimensional space of commensurate and near
commensurate grain boundaries for three-dimensional FCC crystals.

These commensurability questions are not only of practical importance in
efficiently computing the properties of grain boundaries; commensurate 
grain boundaries (especially those with short repeat distances) also have 
especially low energies and high peak stresses~\cite{grain-boundary-geo,wolf-book}.
Modeling the geometry dependence of the peak stress and energy 
could be relatively straightforward if they depended in comprehensible
ways on the five geometrical parameters. The second challenge in 
continuum models of fracture is to {\em incorporate the singularities associated
with commensurate geometries into appropriate functional forms}.

\begin{figure}
\begin{center}
\subfigure[]{\includegraphics[height=5cm]{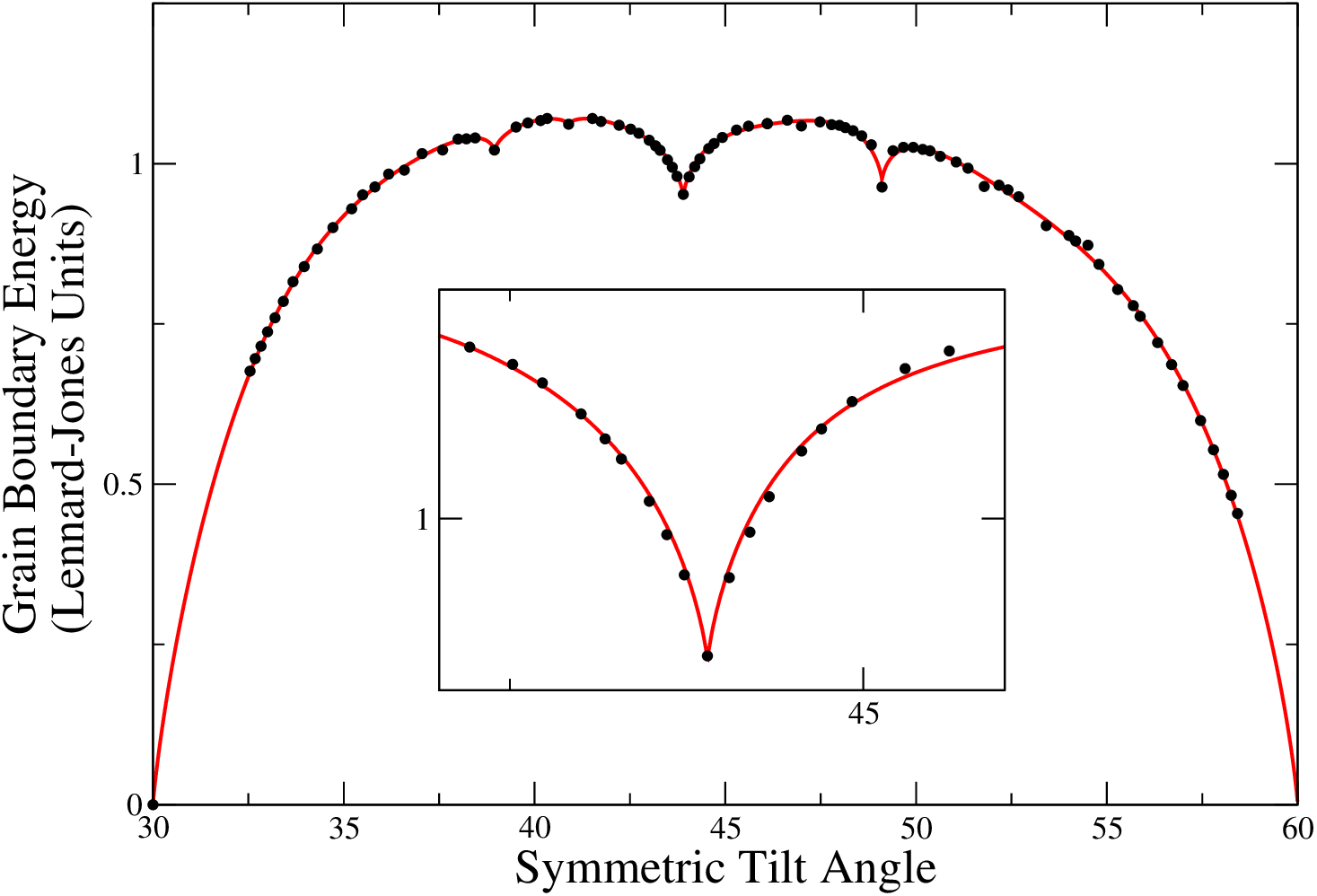}
\label{fig:GBEnergyFit}}
\hskip .2cm
\subfigure[]{\includegraphics[height=5cm]{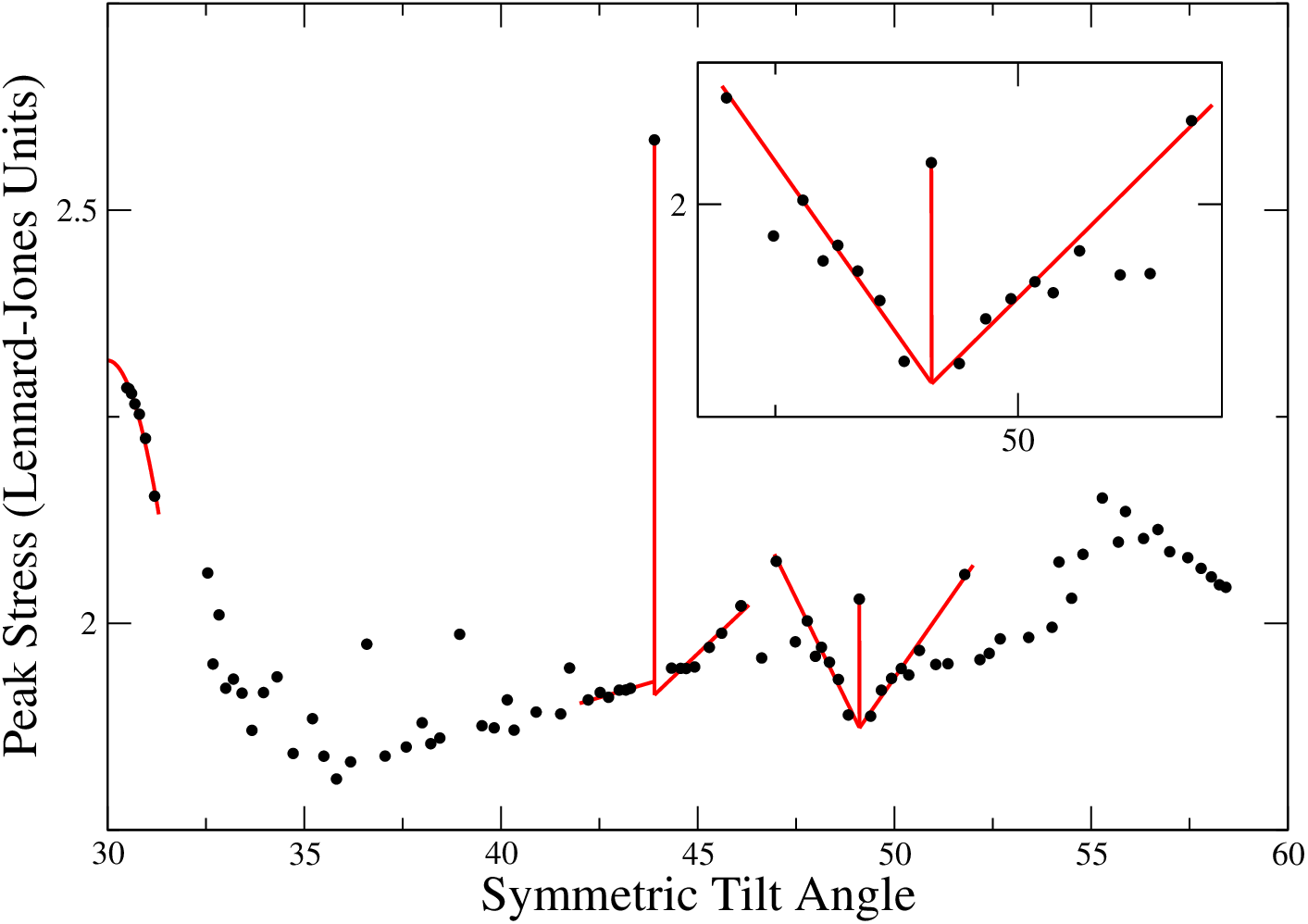}
\label{fig:PeakStress}}
\end{center}
\mycaption[Singularities at Special Grain Boundaries.]{Cusps in energy appear at
high symmetry grain boundaries (figure a) and have the same
$\theta\log\theta$
shape as the energy of low angle grain boundaries. The red line is the functional
form described in~\cite{grain-boundary-geo}.
The peak stress as a function of tilt angle (figure b) is
discontinuous everywhere, with higher values at special tilt angles
corresponding to high symmetry grain boundary geometries.  The dependence of
peak stress on angle near the high symmetry grain boundaries (lines
and parabolas shown) can be explained using the interactions of the extra
dislocations added~\cite{grain-boundary-geo}. The inset
shows the peak stress for the grain boundary with tilt angle 49.1 and the
nearby geometries.}
\end{figure}

It is well established that cusp singularities in the energy occur at 
special high-symmetry grain boundaries with low repeat
distances~\cite{grain-boundary-geo,
sansoz-molinari-structure, chen-GB, wolf-book, old-gb-book}.  The cusps in
energy can be understood by thinking of a high-symmetry boundary as an 
undeformed reference crystal~\cite{grain-boundary-geo}.
Nearby grain boundaries (whose crystallites are rotated by a small angle
$\theta$ from the high-symmetry boundary) are thus described by decorating
the high-symmetry boundary with a few extra dislocations, just as a 
low-angle grain boundary in a crystal can be described as an array of
well-separated dislocations. This analogy leads to a functional form for
grain boundary energy as a function
of tilt angle in which the cusps around the special high symmetry grain
boundaries have the same $\theta \log \theta$ form as low angle grain
boundaries~\cite{grain-boundary-geo}.  Fig.~\ref{fig:GBEnergyFit} shows the
results for a systematic study of symmetric 2D grain boundaries and the
resulting fitting function~\cite{grain-boundary-geo}.

For the peak stress, it is known that there are jumps at the same
special grain boundaries~\cite{chen-GB, wolf-book, grain-boundary-geo}.  
We can also understand these jumps by using the dislocation picture described
above~\cite{grain-boundary-geo}.
As we add a
dislocation to the high symmetry grain boundary, we add a nucleation point for
fracture, causing a discontinuity in the peak stress. As a result, the plot of
peak stress vs. tilt angles is discontinuous at every commensurate geometry
(Fig.~\ref{fig:PeakStress}). By considering the elastic interaction between
the extra dislocations, we have been able to understand also the dependence
of the peak stress in the vicinity of the high-symmetry
boundaries (see Fig.~\ref{fig:PeakStress} and~\cite{grain-boundary-geo}).

\begin{figure}[thb]
\begin{center}
\subfigure[Atomistic simulation at 10\%
strain]{\includegraphics[width=4cm]{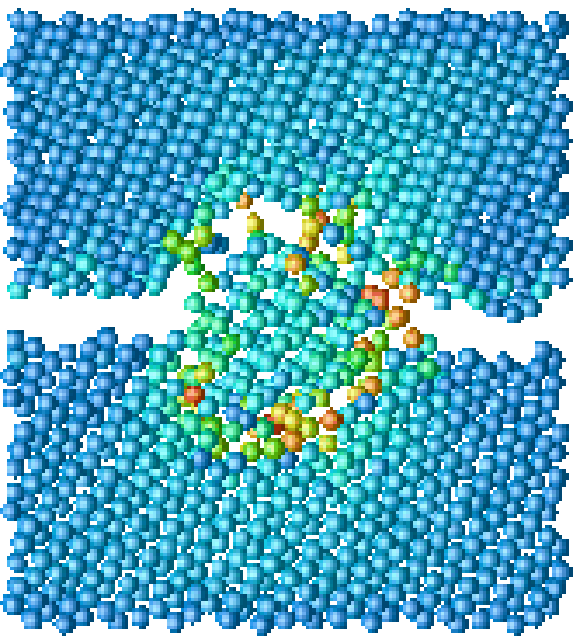}}
\hskip .4cm
\subfigure[FEM simulation at 13.1\%
strain]{\includegraphics[width=4cm]{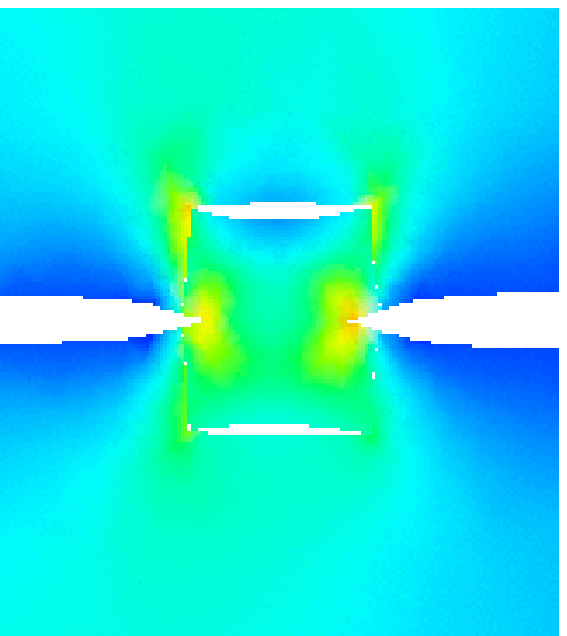}}

\includegraphics{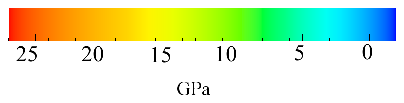}
\end{center}
\mycaption[Comparison of Atomistic and Continuum Fracture Simulations. ]{The figures above show a center cross-section of the atomistic (figure
a) and continuum (figure b) polycrystal simulations.  The geometry is that of a
cube in cube with three crystal orientations for the upper half of the outer
cube, lower half of the outer cube, and the inner cube. An upwards displacement is
imposed on the top face.  The color scale indicates the vertical component of
stress. In the FEM simulation (figure b), we include an interface through the
center of the inner cube to allow for intragranular fracture. The atomistic
simulation fails at a strain 0.02 smaller than the FEM simulation (a
percentage error of 20\%, with the crack propagating partially
intragranularly through the inner cube and partially through the interface at
the top of the inner cube.  The continuum simulation cracks straight through the
center plane of the inner cube.~\cite{cube-in-cube}}
\label{fig:3DAtomVsContinuum}
\end{figure}

Are cohesive laws enough? We have studied this question
computationally~\cite{cube-in-cube} by comparing
a direct atomistic simulation of polycrystalline fracture with a finite-element
simulation of the same geometry using cohesive-law parameters derived
from the same interatomic potential. Fig.~\ref{fig:3DAtomVsContinuum}
shows a snapshot of the two simulations of polycrystalline fracture in
Stillinger-Weber silicon.
Both in this case and for other simulations, the atomistic simulations fail
at significantly lower stresses
than the continuum simulations. Crack nucleation in both atomistic
and continuum simulations happens not in the middle of grain boundaries,
but at triple junction lines, edges, and corners which are not quantitatively
described by the cohesive laws for the grain boundary interfaces.
Similarly, quantitative understanding of how the crack turns,
branches, or goes intragranular (Fig.~\ref{fig:3DAtomVsContinuum})
at triple junctions demands that we understand the effects of the
irregular atomistic configurations at these junctions.

The third challenge is thus to {\em develop an effective computational method
for modeling more complex local geometries}. One in principle could incorporate
an analytical understanding of these local geometries into, for example,
cohesive laws for triple junctions, but the geometrical complexity would
seem daunting. Can one rely here on direct atomistic simulations? Brittle crack
nucleation is a local phenomenon, and the intersection of a growing crack
with a triple junction edge again will generically happen at a point. 
A feasible atomistic simulation of the local region of interest could be
launched whenever the continuum simulation reached a stress state where its
cohesive laws become unreliable. The information about the local geometry
(elastic strains, grain orientations, and impinging crack surfaces)
would be transferred from the continuum simulation to generate the 
atomistic configuration, and the results of the atomistic simulation
(nucleation thresholds, crack branching and turning events) passed back
to the finite-element simulation.

\begin{figure}[thb]
\begin{center}
\subfigure[Initial configuration]{\includegraphics[height=3cm]{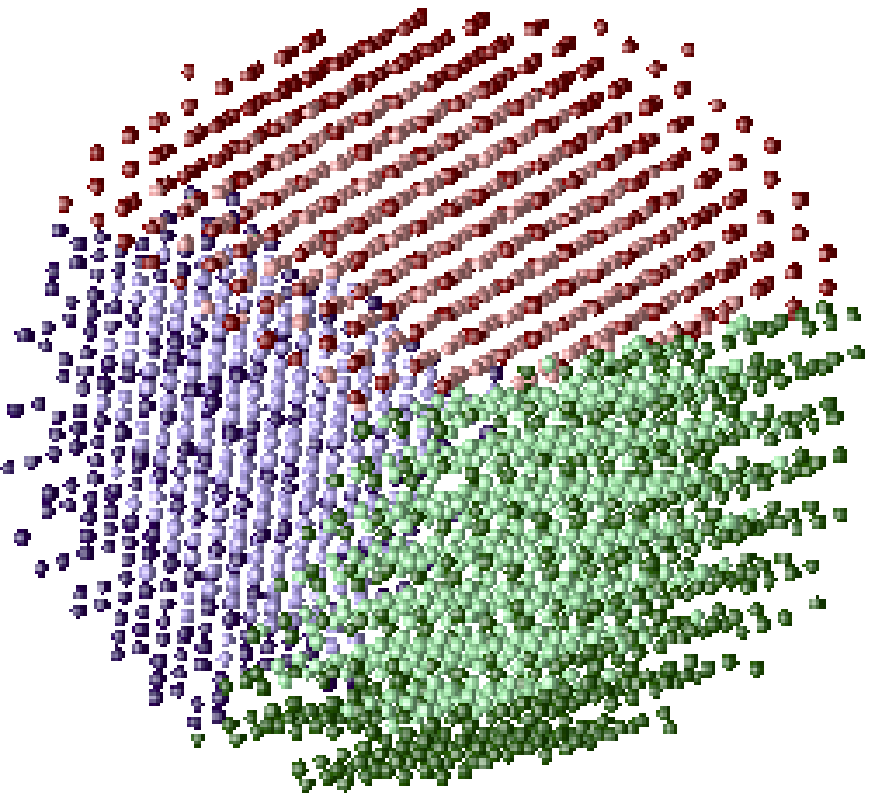}}
\hskip .2cm
\subfigure[Cracked configuration]{\includegraphics[height=3cm]{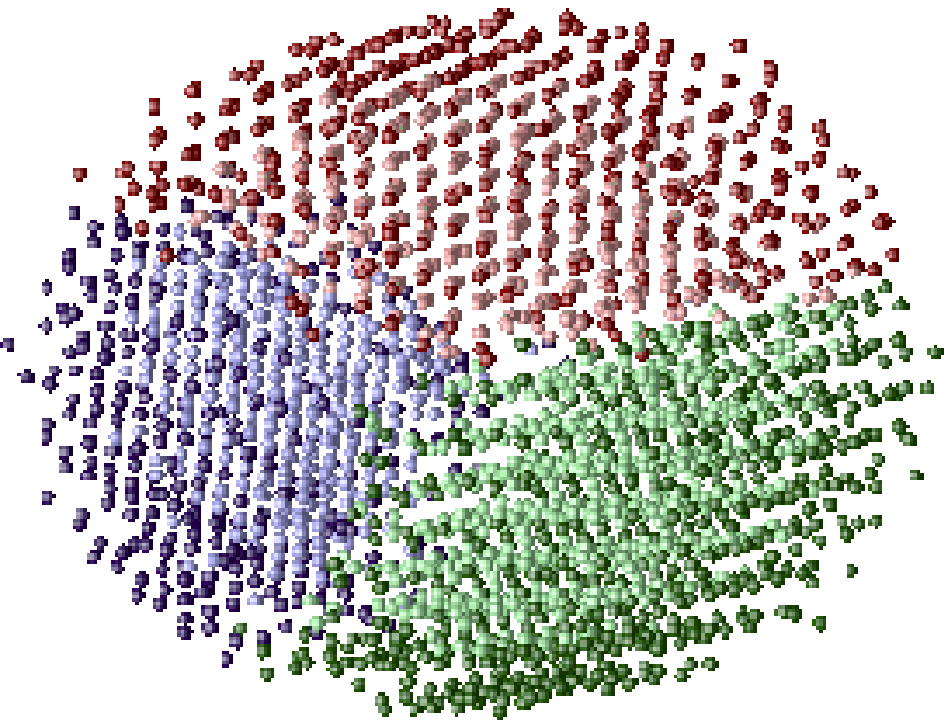}}
\end{center}
\mycaption[Direct Atomistic Simulation of Triple Junctions.]{A cylinder of atoms decorates a finite element edge that coincides with a triple
junction of grains.  Figure a shows the initial configuration.  Figure b shows
the atoms after they have been deformed according to the finite element
displacements and relaxed, opening a crack along two of the interfaces.}
\label{fig:OFEMD}
\end{figure}

As an example of such a method, Fig.~\ref{fig:OFEMD} shows an Overlapping
Finite Elements and Molecular Dynamics (OFEMD) simulation of fracture at a
triple-junction, generated automatically in this
fashion~\cite{cube-in-cube}. The grains were generated using geometry,
orientation, and boundary strains passed from a finite element simulation,
running on a separate machine and communicating either through a command-line
interface, a Web service, or a database. OFEMD uses the
DigitalMaterial~\cite{DigitalMaterial} atomistic simulation environment to
deform and relax the atomistic coordinates, allowing the failure information to
be recorded. OFEMD can be downloaded from~\cite{url:OFEMD}.

We have discussed three main challenges involved in continuum modeling of
polycrystal fracture.  First, exploring the cohesive properties of 3D grain
boundaries involves exploring a 5D space. Second, the peak stress and energy
have singularities at all commensurate grain boundaries. Even if it weren't for
the first two challenges, our comparisons of atomistic and finite element
simulations of polycrystal fracture show that cohesive properties of the
interfaces alone are not enough to model the fracture of polycrystals using
continuum methods.  Sites such as triple junctions, edges, and corners of grains
are important nucleation sites.  In order to resolve this last challenge, we
suggest the use of direct atomistic modeling of local regions of interest.

\begin{acknowledgements}
This work was supported by NSF Grants No. ITR/ASP ACI0085969 and
No. DMR-0218475.  We also wish to thank Drew Dolgert, Surachute Limkumnerd,
Chris Myers, and Paul Wawrzynek.
\end{acknowledgements}

\section*{Competing financial Interests}
The authors declare no competing financial interests.

\bibliographystyle{nature}

\end{document}